\documentclass[10pt]{article}
\usepackage{authblk}
\usepackage[letterpaper, margin=1in, top=1in]{geometry}
\renewenvironment{abstract}{\par\noindent\textbf{\abstractname.}\ \ignorespaces}{\par\medskip}
\usepackage{titlesec}
\usepackage{hyperref}
\usepackage{newtxtext}
\usepackage{newtxmath}
\usepackage{caption}
\usepackage{subcaption}
\usepackage[superscript]{cite}
\usepackage{parskip}
\usepackage{graphicx}
\usepackage{multirow}
\usepackage[table,xcdraw]{xcolor}
\usepackage[labelfont=bf]{caption}
\usepackage{physics}
\usepackage{wrapfig}
\usepackage{tikz}
\usepackage{cleveref}
\usetikzlibrary{positioning}
\usetikzlibrary{shapes}
\titleformat*{\section}{\fontsize{12}{12}\bfseries}
\titlespacing{\section}{0pt}{\parskip}{\parskip}
\titlespacing{\subsection}{0pt}{\parskip}{\parskip}
\titleformat{\subsection}[runin]{\normalfont\normalsize\bfseries}{\thesubsubsection}{10pt}{}
\makeatletter
\patchcmd{\@maketitle}{\LARGE}{\fontsize{14}{17}\selectfont}{}{}
\makeatother

\DeclareCaptionFormat{myformat}{#1#2#3\hrulefill}

\title{\vspace{-6ex}\textbf{Cryptogenic Stroke and Migraine: \\Using Probabilistic Independence and Machine Learning to Uncover Latent Sources of Disease from the Electronic Health Record}\vspace{-1ex}}

\author[1]{Joshua W. Betts}
\author[2]{John M. Still}
\author[2]{Thomas A. Lasko, MD, PhD\vspace{-1ex}}
\affil[1]{Vanderbilt University School of Medicine, Nashville, TN}
\affil[2]{Vanderbilt University Medical Center, Dept. of Biomedical Informatics, Nashville, TN\vspace{-2ex}}
\date{}
\begin{document}

\maketitle


\begin{abstract}
	Migraine is a common but complex neurological disorder that doubles the lifetime risk of cryptogenic stroke (CS).
	However, this relationship remains poorly characterized, and few clinical guidelines exist to reduce this associated risk. 
	We therefore propose a data-driven approach to extract probabilistically-independent sources from electronic health record (EHR) data and create a 10-year risk-predictive model for CS in migraine patients.
	These sources represent external latent variables acting on the causal graph constructed from the EHR data and approximate root causes of CS in our population.		
	A random forest model trained on patient expressions of these sources demonstrated good accuracy (ROC 0.771) and identified the top 10 most predictive sources of CS in migraine patients. 
	These sources revealed that pharmacologic interventions were the most important factor in minimizing CS risk in our population and identified a factor related to allergic rhinitis as a potential causative source of CS in migraine patients. 
\end{abstract}

\section*{Introduction}

Ischemic stroke (IS) is consistently among the top five most common causes of death both in the United States\cite{cdc} as well as worldwide\cite{who} and is a source of significant health-care related costs in developed countries\cite{econ}. Every year, significant resources are spent in the prevention of the "big three" causes of IS\textemdash large vessel atherosclerotic, small-vessel occlusive (lacunar), and cardioembolic \cite{toast,etiology}. However, much less light has been shed on risk-reduction of cryptogenic stroke (CS), which by some estimates account for 30\%-40\% of IS cases in the U.S\cite{cs}. While the simplest definition of CS includes any ischemic stroke that is not determined to be one of the three aforementioned etiologies\cite{asco} (the definition we use in this paper), it should be noted that terminology surrounding CS varies. Some sources use the label of CS to refer only to strokes of undetermined etiology,\cite{toastsss,strokeclassifier} while others also include strokes of determined but uncommon etiologies\cite{ESUS} and/or embolic strokes of undetermined source (ESUS)\cite{esus2} under the cryptogenic umbrella.
 
Despite CS's relative lack of characterization, a link between migraine and CS has been frequently described in the literature \cite{migraineandstroke}. Migraine is  exceedingly common, affecting 10-20\% of US residents,\cite{migraine} and increases the lifetime CS risk by a factor of two. \cite{migraineandstroke} However, few clinical recommendations exist in the literature to reduce this associated stroke risk in migraine patients. These range from the well-established guideline of avoiding estrogen-containing contraception in patients who experience migraine with aura\cite{ACOG,ocp} to identification and closure of a patent foramen ovale \cite{pfoclose,pfocs}\textemdash a common procedure that has become controversial in its off-label use in addressing migraine symptoms\cite{pforeview}. 

Given the poor outcomes associated with stroke and the high prevalence of migraine, identification of additional interventions to reduce the risk of CS in migraine presents a cost-efficient opportunity to reduce overall population morbidity and mortality. The electronic health record (EHR) has long been considered as a potential source of data for investigations into such associations, but has historically been considered too incomplete, noisy, and massive to analyze with typical methods\cite{lasko2013, unsupervised}. Methods utilizing machine learning (ML) have already shown success in creating models from analyzing messy, irregular EHR data\cite{lasko2013, gboost, unsupervised, marco}, though these have historically been tuned for predictive accuracy \cite{predictbreast, predictcancer} as opposed to causal analysis\cite{laskonpj}, limiting interpretability. So while current methods of analysis may yield highly accurate predictive models, they tend to not offer insight into the driving factors behind the disease itself\cite{laskonpj}. However, recent studies have demonstrated the ability of probabilistic independence-based algorithms\cite{lasko, marco} to create models that successfully identify patient-specific root causes of disease\cite{strobl} with only a mild impact on predictive accuracy\cite{marco}. These causative models exhibit enhanced interpretability and allow for identification of complex diseases with heterogenous presentations and identification of their driving factors\cite{laskonpj, ai}.
 
In this study, we make the following contributions. First, we use probabilistic independence-based techniques to disentangle 2000 latent sources of disease and their clinical signatures from the real-world EHR data of a broad sample of neurology patients. Second, we project the EHR data of a large sample of migraine patients onto this signature space and train a causal model that predicts a patient's 10-year risk of CS from these projections. Third, we evaluate this model using a held-out test set to assess accuracy and compare it to alternative models. Finally, we identify which sources have the largest causal effects on CS in migraine patients and may serve as targets for future investigation.

\section*{Methods}

\begin{wraptable}{R}{.44\linewidth}
\captionsetup{format=myformat}
\vspace*{-\baselineskip}
\resizebox{\linewidth}{!}{
\begin{tabular}{|cc|c|cc|}
\hline 
\multicolumn{2}{|c|}{}                                                                                                                                      &                                                                            & \multicolumn{2}{c|}{\begin{tabular}[c]{@{}c@{}}Evaluation Set\\ \small{(Migraine patients)}\end{tabular}}                                            \\ \cline{4-5} 
\multicolumn{2}{|c|}{\multirow{-2}{*}{\textit{Demographic}}}                                                                                                & \multirow{-2}{*}{\begin{tabular}[c]{@{}c@{}}Discovery \\ Set\end{tabular}} & \multicolumn{1}{c|}{\begin{tabular}[c]{@{}c@{}}Negative \\ for CS\end{tabular}} & \begin{tabular}[c]{@{}c@{}}Positive \\ for CS\end{tabular} \\ \hline
\multicolumn{2}{|c|}{\textit{\begin{tabular}[c]{@{}c@{}}N$^\circ$ of unique records\end{tabular}}}                                                          & 309,759                                                                    & \multicolumn{1}{c|}{71,206}                                                     & 1,670                                                      \\ \hline
\multicolumn{1}{|c|}{}                                                                                             & \cellcolor[HTML]{E0E0E0}\textit{Male}  & \cellcolor[HTML]{D0D0D0}46.1\%                                             & \multicolumn{1}{c|}{\cellcolor[HTML]{D0D0D0}23.4\%}                             & \cellcolor[HTML]{D0D0D0}25.5\%                             \\ 
\multicolumn{1}{|c|}{\multirow{-2}{*}{\textit{Sex}}}                                                               & \textit{Female}                        & 53.9\%                                                                     & \multicolumn{1}{c|}{76.6\%}                                                     & 74.4\%                                                     \\ \hline
\multicolumn{1}{|c|}{}                                                                                             & \cellcolor[HTML]{E0E0E0}\textit{White} & \cellcolor[HTML]{D0D0D0}74.9\%                                             & \multicolumn{1}{c|}{\cellcolor[HTML]{D0D0D0}78.0\%}                             & \cellcolor[HTML]{D0D0D0}73.8\%                             \\ 
\multicolumn{1}{|c|}{}                                                                                             & \textit{Black}                         & 12.2\%                                                                     & \multicolumn{1}{c|}{10.6\%}                                                     & 16.5\%                                                     \\  
\multicolumn{1}{|c|}{}                                                                                             & \cellcolor[HTML]{E0E0E0}\textit{Other} & \cellcolor[HTML]{D0D0D0}5.2\%                                              & \multicolumn{1}{c|}{\cellcolor[HTML]{D0D0D0}4.9\%}                              & \cellcolor[HTML]{D0D0D0}2.9\%                              \\ 
\multicolumn{1}{|c|}{\multirow{-4}{*}{\textit{Race}}}                                                              & \textit{Unknown}                       & 9.0\%                                                                      & \multicolumn{1}{c|}{7.8\%}                                                      & 3.1\%                                                      \\ \hline
\multicolumn{1}{|c|}{}                                                                                             & \cellcolor[HTML]{E0E0E0}\textit{Mean}  & \cellcolor[HTML]{D0D0D0}48                                                 & \multicolumn{1}{c|}{\cellcolor[HTML]{D0D0D0}42}                                 & \cellcolor[HTML]{D0D0D0}51                                 \\ 
\multicolumn{1}{|c|}{\multirow{-2}{*}{\textit{\begin{tabular}[c]{@{}c@{}}Age, yrs\end{tabular}}}}             & \textit{IQR}                           & {[}32, 66{]}                                                               & \multicolumn{1}{c|}{{[}28, 57{]}}                                               & {[}40, 67{]}                                               \\ \hline
\multicolumn{1}{|c|}{}                                                                                             & \cellcolor[HTML]{E0E0E0}\textit{Mean}  & \cellcolor[HTML]{D0D0D0}9                                                  & \multicolumn{1}{c|}{\cellcolor[HTML]{D0D0D0}10}                                 & \cellcolor[HTML]{D0D0D0}12                                 \\ 
\multicolumn{1}{|c|}{\multirow{-2}{*}{\textit{\begin{tabular}[c]{@{}c@{}}Record \\ length, yrs\end{tabular}}}} & \textit{IQR}                           & {[}3, 15{]}                                                                & \multicolumn{1}{c|}{{[}3, 16{]}}                                                & {[}4, 20{]}                                                \\ \hline
\end{tabular}}
\caption{Demographic data of study populations and subsets. IQR: interquartile range}
\end{wraptable}

This study was conducted at Vanderbilt University Medical Center (VUMC) and was determined by its Institutional Review Board to be non-human-subjects research (\#171392).

\subsection*{Data collection and set definitions} 

\begin{wrapfigure}{r}{.44\textwidth}
\captionsetup{format=myformat}
\vspace*{-4.4\baselineskip}
\includegraphics[width=.44\textwidth]{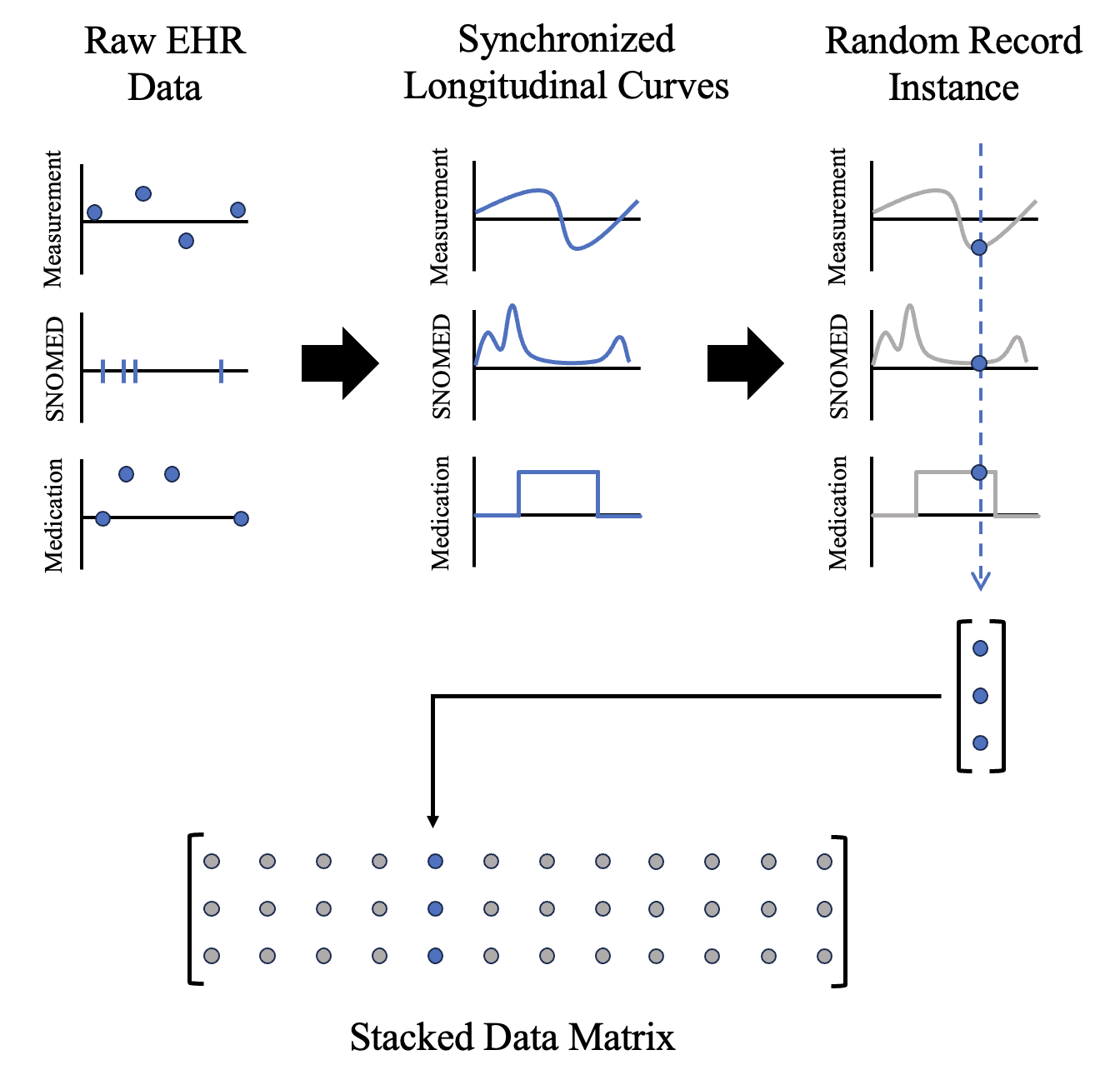}
\caption{Preprocessing pipeline for EHR data. Longitudinal curves are inferred from the raw observational data and stacked, synchronizing the discrete data channels and allowing for all clinical variables to be sampled at any given point in time. Random instances of these curvesets records are then sampled and concatenated into the stacked data matrix $\textbf X$.}
\label{figure1}
\vspace*{-\baselineskip}
\end{wrapfigure}

All data was extracted from the Synthetic Derivative (SD) \cite{sd}, VUMC's de-identified EHR mirror of over 3 million patient records from 2004-present. With a cutoff date of 08/31/24, we collected historical EHR data across the 9000 most common quantitative clinical variables from all 311,429 patient records that contained at least one ICD-10 code in the range of G40-G47, representing a broad range of paroxysmal neurological disease. We then defined our \textit{evaluation set} as the subset of these records containing an ICD-10 code of G43, corresponding to any migraine diagnosis ($N = 72,876$). Each patient ID in the evaluation set was then randomly assigned to either the \textit{training set} ($N = 54,654$) or the \textit{test set} ($N = 18,222$) with a probability of 0.2 of being assigned to the latter. Patient IDs in the test set were then subtracted from our initial population to ensure there was no data leak from records in \hspace{17cm} the test during model training.  This finalizes our \textit {discovery set} \hspace{10cm}\\($N = 309,759$), with demographics detailed in \textbf{Table 1}.

\subsection*{Data pre-processing and continuation}

Instances of a patient's health data occur at discrete, asynchronous points in the record, with large gaps existing between encounters with the healthcare system. Sampling a record at a random point is thus unlikely to give values for a majority of chart variables. This noisy, discrete data  must be transformed into continuous and synchronous data that can be sampled at any point in time. We followed the methods employed by Lasko et al.\cite{lasko} and Mota et al.\cite{marco} for data preprocessing. In summary, we computed longitudinal curves for all 9000 chart variables for each patient with a temporal resolution of one day from the discrete  and asynchronous EHR observations (\textbf{Fig.1}). The process used for curve generation for each variable depended on its data modality, as follows. 

For medical conditions, represented by SNOMED billing codes in the SD, we constructed smooth code-intensity curves from discrete instances using a variation of the Random Average Shifted Histograms (RASH) algorithm\cite{RASH}. In the event of missing data, we imputed a constant curve with baseline annual event-arrival probability of once per 20 years. For clinical measurements such as vitals and lab values, we used the Piecewise Cubic Hermite Interpolating Polynomial (PCHIP)\cite{pchip} to generate smooth continuous curves between the recorded discrete values. Records without observations for a given variable were assigned a constant curve of the population median. For medications, we opted to create binary curves that approximate medication adherence, given that medication data in the SD solely indicates presence on a visit's medication list. We assumed continuation of regimens noted on contiguous visits, imputing a value of 1. If a subsequent visit's list dropped a medication, we assumed discontinuation at the temporal midpoint. Medications that were never noted on a given chart were assigned the zero function. Finally, demographics such as race and sex were represented by constant binary variables based on the record's most recent value.

We stacked each curve longitudinally into an $m \times t$ curveset for each patient, where $m$ is the number of clinical variables and $t$ is the number of days spanned by the record. We then sampled $n$ curveset cross sections at uniformly random times with an average density of one sample per record-year. This results in longer charts having a higher probability of being sampled more than once, while some shorter charts were not sampled at all. 

Merging all cross sections yields a dense data matrix $\textbf{X} \in \mathbb R ^{m \times n}$ where each column $\vb x_{j}$ of $\textbf{X}$ represents a single estimate of a complete patient state taken at a specific time in a patients record across all $m = 9000$ extracted clinical variables. Our matrix $\textbf X$ was then standardized by row to bring all variables into roughly the same scale; measurements, log-transformed condition code intensities, and ages were centered so that their means were zero and scaled by two standard deviations\cite{std} while binary variables were left untouched. 

\subsection*{Clinical signature discovery} 

To disentangle the clinical signatures from our stacked patient data matrix $\textbf X$, we relied on the principles of probabilistic independence. This approach stems from the assumption that latent sources of disease act independently from each other and leave a unique signature on the clinical history of the patient which is then recorded, albeit imperfectly, into the EHR.\cite{lasko,marco,strobl} Disentanglement was achieved via unsupervised Independent Component Analysis (ICA) \cite{ica} decomposition of our data matrix $\ \textbf{X} = \textbf{AS}\ $ using our lab's custom-optimized implementation\cite{stillica} of the \texttt{FastICA} algorithm\cite{fastica}, where $\textbf S \in \mathbb R ^{k \times n}$ is the source matrix and $\textbf{A} \in \mathbb R ^{m \times k}$ is the mixing matrix which encodes how our disease sources act together to produce the observed EHR data represented in $\textbf{X}$\cite{lasko}.

The \texttt{FastICA} algorithm ensures the rows of $\textbf S$ are (maximally) mutually independent; thus each row $\vb s'_i$ represents the expression level of a disentangled latent source of disease. Each column $\vb s_j$ of $\textbf S$ then quantifies the extent to which each source $j$ is expressed in patient record instance $\vb x_j$.
Meanwhile, each column $\vb a_j$ of $\textbf{A}$ represents the clinical signature of source $j$ \textemdash i.e. the $m$ linear mixing weights that describe the unique imprint of changes to the EHR data by its corresponding latent source $\vb s'_j$. Note that $k$, the number of sources inferred by our ICA decomposition, would ideally be close to the intrinsic dimensionality of $\textbf{X}$ to avoid over- or under-learning \cite{ica}; however, we were limited to a value of $k = 2000$ due to memory constraints. 

\subsection*{Record labeling}

\begin{wrapfigure}{r}{.36\textwidth}
\vspace*{-.2\baselineskip}
\captionsetup{margin=.2cm}
\resizebox{\linewidth}{!}{
\begin{tikzpicture}%
  [data/.style=
    {draw,minimum height=0.7cm,minimum width=2cm,align=center},
   filter/.style=
    {draw,minimum height=1.0cm,minimum width=2.5cm,align=center},
   database/.style=
    {draw,minimum height=1.8cm,minimum width=3cm,align=center},
   flow/.style={thick,-stealth},
   apply/.style={}
  ]
  \node[database] (e0) {\textbf{Evaluation Set}\\\small(Migraine patients)\\$N = 72,876$};
  \node[data,below=of e0] (e1) {$N=3991$};
  \node[data,below=of e1] (e2) {$N=2802$};
  \node[data,below=of e2] (e3) {$N=1939$};
  \node[data,below=of e3] (e4) {\small \textbf{Positive for CS}\\$N=1670$};
  \draw[flow] (e0) -- coordinate(e0e1) (e1);
  \draw[flow] (e1) -- coordinate(e1e2) (e2);
  \draw[flow] (e2) -- coordinate(e2e3) (e3);
  \draw[flow] (e3) -- coordinate(e3e4) (e4);

  \node[right=of e0] (excl) {\ \ \ \ \ \ \ \ \ \ \ \ \ \ \ \ \  };
  \node[filter] (f1) at (e0e1-|excl) {\small No stroke code\\($N=68,885$)};
  \draw[flow] (e0e1) -- (f1);
  \node[filter] (f2) at (e1e2-|excl) {\small Stroke codes not \\ \small specific for CS\\($N=1189$)};
  \draw[flow] (e1e2) -- (f2);
    \node[filter] (f3) at (e2e3-|excl) {\small Co-occuring \\ \small non-cryptogenic\\ \small stroke code\\($N=863$)};
  \draw[flow] (e2e3) -- (f3);
      \node[filter] (f4) at (e3e4-|excl) {\small Single stroke code\\($N=269$)};
  \draw[flow] (e3e4) -- (f4);
 
\end{tikzpicture}}
\captionsetup{format=myformat}
\caption{Flowchart detailing the multi-step inclusion and exclusion criteria (based on ICD-10 diagnosis) codes used to label patient records as either positive or negative for CS.} 
\vspace*{-\baselineskip}
\end{wrapfigure}

Due to CS's varying definitions and status as a diagnosis of exclusion, there is no single ICD-10 code or even set of codes that neatly correspond to a diagnosis of CS. Additionally, some non-specific codes may apply to both cryptogenic and non-cryptogenic strokes, and clinicians may sometimes use inaccurate billing codes when documenting a stroke.
To account for these complexities, we developed a 4-step algorithm of inclusion and exclusion criteria in order to label records in our evaluation set as positive ($N = 1670$) or negative ($N = 71,206$) for CS (\textbf{Fig. 2}).  

First, the set of ICD-10 stroke codes were identified that could reasonably be applied to a CS event, forming the initial set of inclusion criteria (G43.6, I63, I67.8). We then evaluated these codes for how often they coincided with other ICD-10 codes for IS that are specific for a non-cryptogenic etiology in the same patient chart. Codes that had more than a 30\% coincidence rate were considered overly non-specific for CS and removed as inclusion criteria, yielding a more specific set of inclusion codes (G43.6, I63.212, I63.52, I63.6, I63.8, I63.9, I67.848). 
The same codes that are specific for non-cryptogenic etiologies were then used as exclusion criteria to account for patients who initially received a diagnosis of CS but whose stroke was eventually determined to be of another etiology (I63.0, I63.1, I63.3, I63.4). Additionally, chart review revealed that in the majority of records with only one stroke code instance, the code was either applied erroneously or was applied to cases of suspected stroke during workup that were later determined to be negative. Therefore, having an instance of only one stroke code was chosen as a final exclusion criterion. 
Note that the number of records labeled as positive for CS in our evaluation set is roughly 40\% of the number of stroke patients in our evaluation set. This proportion is in line with current literature estimates,\cite{cs,toast} suggesting that these criteria are not grossly over- or under-counting instances of CS. 

\subsection*{Model creation}

\begin{wrapfigure}{l}{.5\textwidth}
\centering
\begin{minipage}{\linewidth}
\centering
\captionsetup[subfigure]{justification=centering}
\begin{tikzpicture}%
  [
   flow/.style={thick,-stealth,shorten >=0.15cm,shorten <=0.15cm},
   apply/.style={}, node distance = .8cm
  ]
  \node[draw,fill = teal!20, draw = teal, ultra thick] (l1) {\textbf{Label}};
  \node[left=of l1, draw,fill = blue!20, draw = blue, circle] (o3) {\textbf{$\ O\ $}}; 
    \draw[flow] (o3) -- coordinate(l1o1) (l1);
    \node[left=of o3, draw = purple, dashed, very thick] (c1) {{Disease}};
    \node[above=of c1, draw = purple, dashed, very thick] (c2) {{Effects}}; 
    \node[below=of c1, draw = purple, dashed, very thick] (c3) {{Causes}};   
    \draw[flow] (c3) -- coordinate(l1o1) (c1);
    \draw[flow] (c1) -- coordinate(l1o1) (c2);     
    \draw[flow] (c1) -- coordinate(l1o1) (o3); 
    \draw[flow] (c2) -- coordinate(l1o1) (o3); 
    \draw[flow] (c3) -- coordinate(l1o1) (o3);     
    
\end{tikzpicture}
\subcaption{Ideal causal graph where conceptual latent variables are separated into individual nodes. Node $O$ represents a network of many EHR observations which may influence each other.\\ \ }
\end{minipage}%

\begin{minipage}{\linewidth}
\captionsetup[subfigure]{justification=centering}
\centering
\begin{tikzpicture}%
  [
   flow/.style={thick,-stealth,shorten >=0.15cm,shorten <=0.15cm},
   apply/.style={}, node distance = .9cm
  ]
  \node[draw,fill = teal!20, draw = teal, ultra thick] (l1) {\textbf{Label}};
  
   \node[below right =of l1, draw,fill = blue!20, draw = blue, circle] (o2) {\textbf{$O_{l}$}};
    \draw[flow] (o2) -- coordinate(l1o1) (l1);
    \node[left=of o2, draw,fill = blue!20, draw = blue, circle] (o1) {\textbf{$O_k$}};
    \draw[flow] (o1) -- coordinate(l1o1) (l1);
    \draw[flow] (o1) -- coordinate(o1o2) (o2);
    \node[left=of o1, draw,fill = blue!20, draw = blue, circle] (o3) {\textbf{$O_j$}}; 
    \draw[flow] (o3) -- coordinate(l1o1) (l1);
    \draw[flow] (o3) -- coordinate(o3o1) (o1);
    \node[below=of o1o2, draw = purple, very thick, dashed, circle] (e1) {\textbf{$E_l$}};
    \draw[flow] (e1) -- coordinate(l1o1) (o2);
    \node[below=of o3o1, draw = purple, very thick, dashed, circle] (e2) {\textbf{$E_k$}};
    \draw[flow] (e2) -- coordinate(l1o1) (o1);
    \node[left=of e2, draw = purple, very thick, dashed, circle] (e3) {\textbf{$E_j$}};
    \draw[flow] (e3) -- coordinate(l1o1) (o3);
    \node[above left =of o3, draw,fill = blue!20, draw = blue, circle] (o4) {\textbf{$O_i$}}; 
    \node[above left =of e3, draw = purple, very thick, dashed, circle] (e4){\textbf{$E_j$}};
    \draw[flow] (e4) -- coordinate(l1o1) (o4);
    \draw[flow] (o4) -- coordinate(o3o1) (o1);
    \draw[flow] (o4) -- coordinate(o3o1) (l1);
    \draw[flow] (o4) -- coordinate(o3o1) (o3);
    
\end{tikzpicture}
\subcaption{Causal graph where the observation network $O$ from \textbf{Fig.3(a) }is expanded into an arbitrary example arrangement of 4 observation nodes, each with its own error term identifiable by ICA. These error terms are the root causes of our label to the degree that the data permits for identification.}
\end{minipage}
\captionsetup{format=myformat}
\caption{Two views of the structural causal graph for an arbitrary patient, adapted from Mota et al\cite{marco}. Nodes represent the model's variables while edges (arrows) represent unidirectional and transitive causal relationships. The green rectangular node corresponds to the binary yes/no CS label; shaded blue nodes correspond to observations in the EHR; dashed red circles are latent nodes.}
\label{Figure 3}
\vspace*{-\baselineskip}
\end{wrapfigure} 

The longitudinal curves for all patient charts in the evaluation set were sampled (\textbf{Fig.1}) with an average density of one sample per record-year within a 10-year sampling window. For patient records without any instance of a stroke diagnosis code, the end of this window coincides with the end of the record. For patient records with a stroke code, the window ends one month prior to the earliest occurrence of a stroke code, leaving a buffer such that data from the encounter where the stroke was diagnosed is excluded but ensuring signatures are prospective relative to the stroke occurrence. This results in a prediction time horizon of one month to 121 months; in other words, the model makes a binary prediction whether a patient will have a cryptogenic stroke within the next 10 years.
 
Just as was done with the discovery set data, each of these synchronized patient chart instances $\vb e_j$ were then stacked longitudinally into a dense data matrix $\textbf E$. This matrix was then projected onto the clinical signature space by multiplying by the inverse of the mixing matrix discovered by our earlier ICA decomposition: $\textbf{S} = \textbf{A}^{-1}\textbf{E}$.
Each pair of corresponding columns $\vb e_j$ in $\textbf E$ and $\vb s_j$ in $\textbf S$ are therefore representations of the same information but in two different vector bases, with the latter encoding each instance of patient data how strongly it expresses each of the 2000 learned signatures. 
We then used these projections, along with their corresponding labels (\textbf{Fig. 2}) to train a non-linear random forest (RF) model (\texttt{scikit-learn} 1.6.0, Python 3.10.16) to predict 10-year cryptogenic stroke risk. Although alternative architectures such as XGBoost are generally more accurate when making predictive models, \cite{gboost} RF models have shown superior accuracy and stability in estimating causal effects \cite{lasko, marco}, leading to this study's choice of model architecture.  
The \texttt{optuna} package was used to search for the optimal model hyperparameters with the 6-fold cross-validation area under the receiver operating characteristic curve (AUROC) as the primary endpoint being optimized on our training set data. The hyperparemeter space was searched using a combination of random search, grid search, Bayesian optimization, and human-guided search. Once an optimal set of hyperparameters were identified, a final RF model was trained using all projections from the training set as inputs and evaluated on projections from the test set, using the AUROC as the primary measure of accuracy. 

Once the primary RF model had been trained and evaluated, we trained another secondary model using the same methods as above as a comparator for our primary model. This model was trained on re-sampled instances from the evaluation set and used alternative positive labels indicating general ischemic stroke as inputs. The resulting model was then trained to predict the 10 year likelihood of general ischemic stroke.  

\subsection*{Latent causes}

The data generation process of this problem of recognizing latent causes of CS in the health record is displayed in \textbf{Fig. 3}. The disease process, along with its causes and effects, are the most upstream latent nodes on the causal graph. These upstream nodes cause a set of observations, $O$, which are recorded into the EHR by the clinician. These observations include diagnosis codes, demographics, laboratory values, and medication records, and they may influence one another. Then, based solely on the diagnosis code observations in a given record, a binary yes/no label for CS can be applied via the process detailed in \textbf{Fig. 2}. 

The true root causes of our final label exist within the parentless "Causes" node in \textbf{Fig. 3(a)}. However, we only have access to imperfect observations of the effects of these causes \cite{laskonpj,strobl}. Under the LiNGAM model assumptions \cite{lingam}, the latent sources whose expressions $\vb s'_j$ are captured by our ICA source matrix $\textbf{S}$ represent unobserved latent causes of the target sink node (in this case, our CS label)\cite{strobl,lasko}. Formally, these disentangled sources correspond to the exogenous independent \textit{error terms} of the structural equation model that we assume to describe the underlying causal process  of our problem (\textbf{Fig. 3(b)}). Since causal relations are transitive, these error terms do in fact exert a causal relationship onto the final label assigned to each record. In summary, these error terms (dashed nodes in \textbf{Fig. 3(b)}) are compact representations of the true unobserved causes inferred using the data available in the EHR; therefore, the data generative process in \textbf{Fig. 3(a)} can be equivalently be represented by \textbf{Fig. 3(b)}\cite{strobl}. 

\subsection*{Feature importance}

To compare the degree of causation each source exerts on CS in our migraine patients, we computed the SHapley Additive exPlanations (SHAP) value\cite{shap} for model features across each record in the test set. Because the features used by our model that predict the probability $P(\text{label}\ |\ \textbf{S})$ are the mutually independent error terms of our structural causal model \textbf{(Fig. 3(b))}, the SHAP value of each input feature is a quantitative estimate of the causal effect of that latent source on the record's final label.\cite{strobl}. If the SHAP value is positive, that source increases the probability of a positive CS label; if negative, the source decreases the probability of a positive CS label. 

\section*{Results}

\subsection*{Model performance}

\begin{wrapfigure}{r}{.30\linewidth}
\vspace*{-\baselineskip}
\captionsetup{format=myformat}
\centering
\includegraphics[width=.9\linewidth]{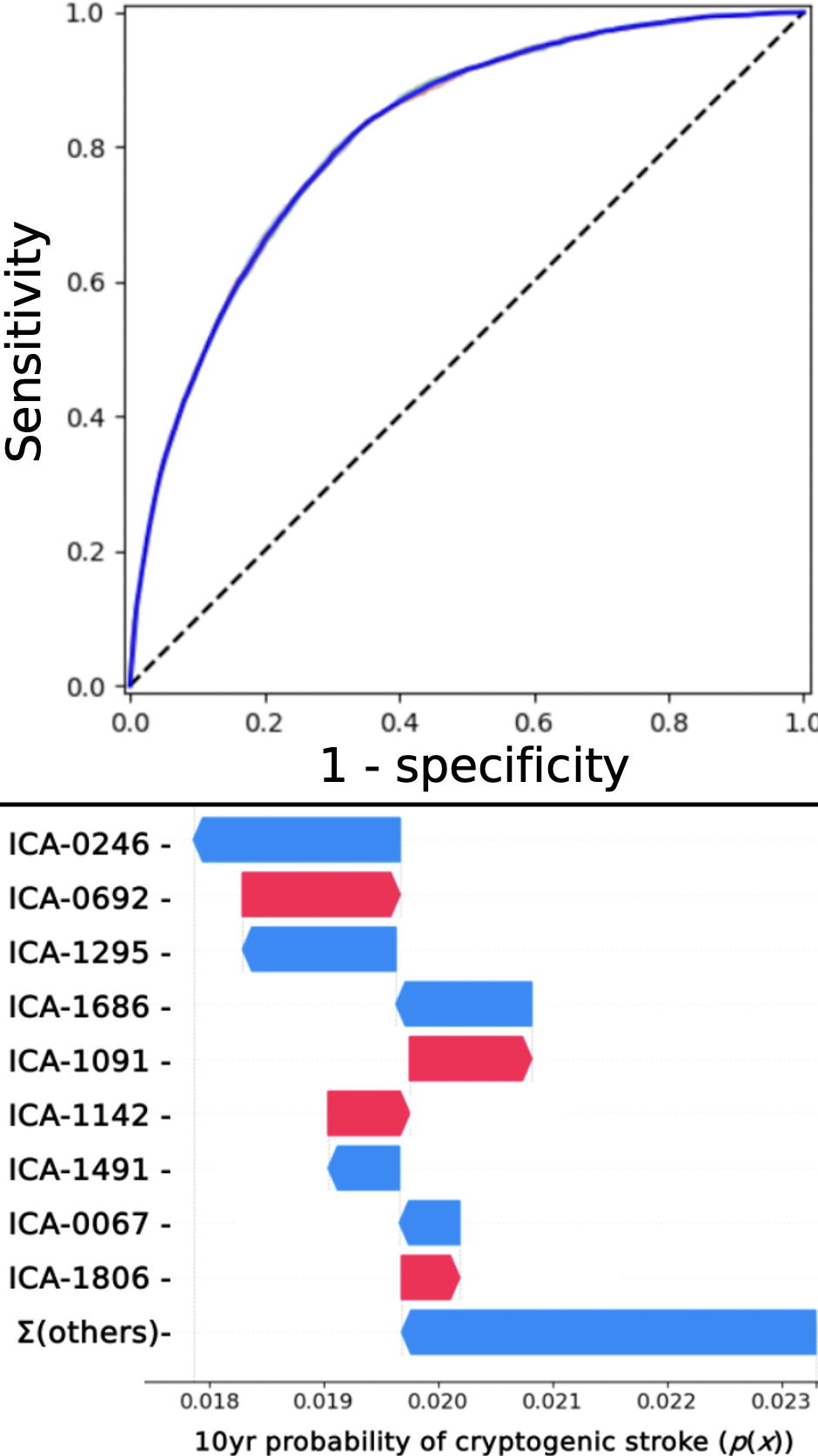}
\caption{\textbf{Above}: ROC curve for test set evaluation (AUC = 0.782).\\\textbf{Below}: Waterfall chart showcasing the model's decision-making for a random patient record, illustrating its ability to provide patient-level causal explanations for a record's final label.}
\label{Figure 4}
\vspace*{-1.8\baselineskip}
\end{wrapfigure}

The mean AUROC of our model when evaluated via 6-fold cross validation with optimal hyperparameters was 0.771. The AUROC when trained on all training set projections and evaluated against the test set was 0.782 (\textbf{Fig.4}). This is lower than the test accuracy of our secondary RF model trained to assess sources of general ischemic stroke, which had a test AUROC of 0.805. The accuracies of both of these models are in line with other published cardiovascular risk estimators that see regular clinical use, such as the ASCVD risk score which has been shown to have an AUROC between 0.65-0.80\cite{ascvd,ascvd2}. 

\subsection*{Inferred causes of CS}

Sources were arranged in descending order by mean absolute SHAP value, with the top 10 most predictive sources of CS shown in \textbf{Fig.5}. The majority of these exhibit a common pattern: the highest-featured (and usually dominant) chart variable is a medication known to reduce stroke risk, and the remaining variables comprise the constellation of clinical diagnoses, symptoms and drugs one would expect to see in a patient taking said medication (\textbf{Fig. 6}). Records with negative expressions for these sources largely had positive SHAP values (indicating increased predicted stroke risk), and vice versa. 

Positive expressions of these sources are easy to interpret\textemdash they indicate that the medication is present in the record, and at least some of the features are as well. Records with a zero or near-zero value for these sources are similarly easy\textemdash neither the medication nor the associated features are present. However, records with negative expressions for these sources are harder to interpret, despite these records yielding the highest positive SHAP values for the corresponding signature. Upon chart review, records that negatively express sources are often partial matches, where some of the variables further down the signature description diagram \textbf{(Fig.6)} are present, while the dominant variable(s) are not. In the case of these medication-dominated sources, this looks like a record that does not contain a mention of the indicated medication but does mention diagnoses associated with said medication.

\begin{figure}[h]
\captionsetup{format=myformat}
  \begin{minipage}[c]{0.64\textwidth}
    \includegraphics[width=\textwidth]{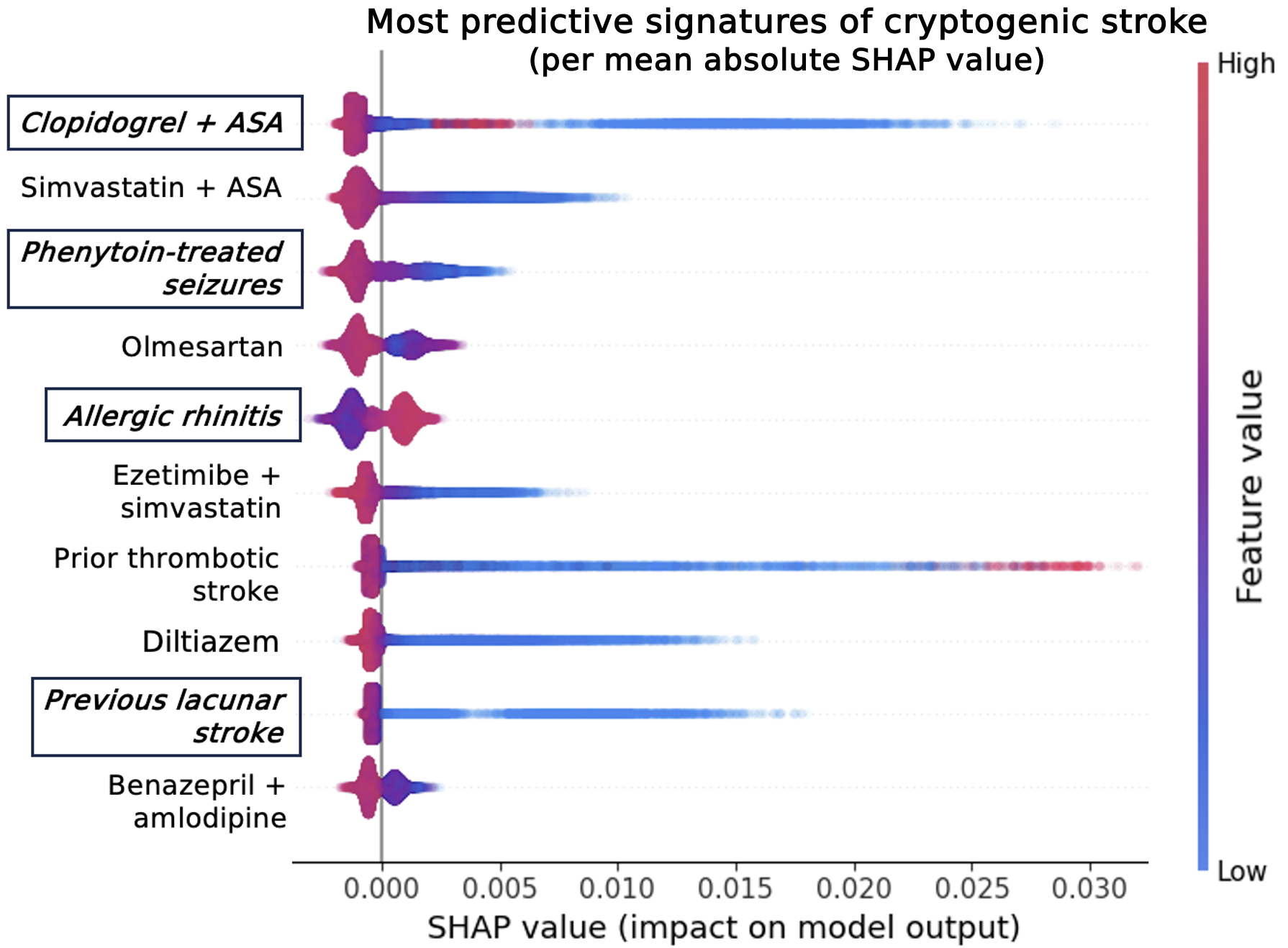}
  \end{minipage}\hfill
  \begin{minipage}[c]{0.34\textwidth}
    \caption{
       Distribution of SHAP values (in log-odds units) for the 10 sources found to be most predictive of CS, color-coded by source expression value. Sources are ranked in descending order by mean $\abs{\text{SHAP}}$. Positive SHAP values indicate increased CS risk; negative SHAP values indicate reduced risk. Signatures with boxed labels are uniquely causative of CS (i.e. did not also rank as a top source for general IS). \\The majority of top sources correspond to common preventative drug regimens and are unsurprisingly negative predictors in the model. However, sources with dominant features of phenytoin and allergic rhinitis were unexpected inclusions among the top sources.
    } \label{Figure 5}
  \end{minipage}
\vspace*{-1.3\baselineskip}
\end{figure}

SHAP values were also computed for the general IS model, and the top features were compared between models. There was significant overlap between top features. However, as many strokes are incorrectly labeled as cryptogenic,\cite{toastsss} this is to be expected, as is it plausible that different stroke types may share common causes. Sources related to lipid-lowering agents and antihypertensives, which have been repeatedly shown to reduce stroke risk across etiologies\cite{zetia,benazepril,amlodipine,statin}, had strong estimated protective effects in both models; meanwhile while signatures that were uniquely causative in the CS model are noted in \textbf{Fig.5}. These include a source primarily characterized by the anti-epileptic phenytoin, which was protective against cryptogenic stroke in the model, and a source related to allergic rhinitis, which had causal effects on CS. The source that was the most uniquely protective of CS corresponded to treatment with clopidogrel; this aligns with current literature and guidelines which recommend anti-platelet therapy for primary and secondary prevention of CS\cite{clopidogrel}.

Additionally, if signatures are ranked by max absolute SHAP value instead of mean absolute SHAP value, a source corresponding to sickle cell disease appears. This high max SHAP value indicates that this source is highly impactful when present in the record. However, the model does not consider lack of expression of this source to be protective, meaning records without a sickle cell diagnosis have a near-zero SHAP value. Since sickle cell disease is a rare diagnosis, this source's mean absolute SHAP value remains low. 

Meanwhile, the source related to allergic rhinitis sees the opposite effect. Its max absolute SHAP value is significantly lower than other top predictors. However, the model is incredibly consistent in how it estimates the effect of this source\textemdash positive expression invariably increases the risk of CS in the model, and its negative expression decreases risk. So although the associated change in CS risk is small (0.001 on average), it is consistently non-zero, resulting in a large mean absolute SHAP value. 

\begin{figure}[h]
\centering
\includegraphics[width=0.9\linewidth]{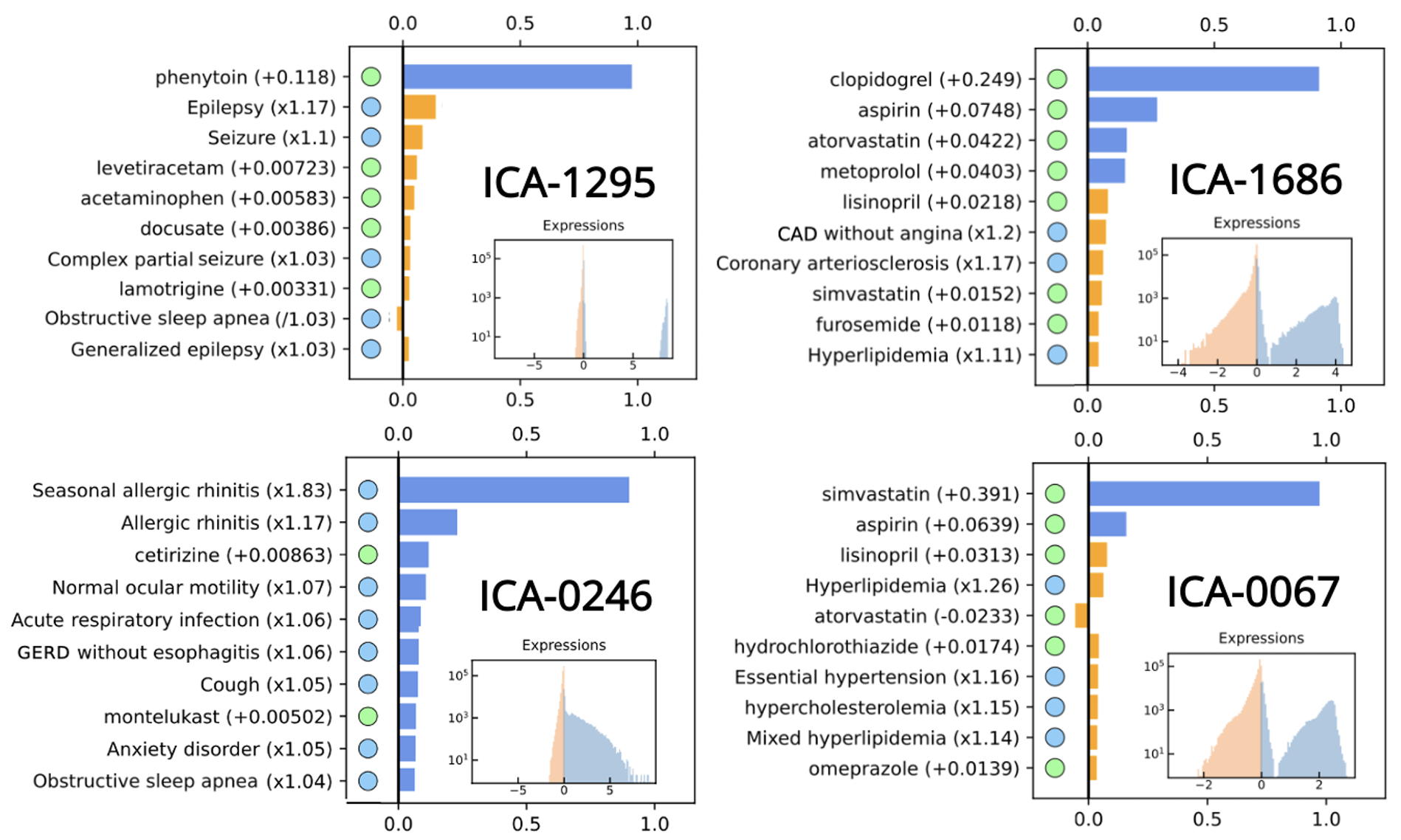}
\captionsetup{format=myformat}
\caption{Signature description diagrams for 4 of the most predictive sources, as shown in \textbf{Fig.5}. Bar length represents the normalized change in a clinical variable's value for each unit of expression, shown in parenthesis. Inset log-scale histograms show source expression levels across all discovery set instances.}
\vspace*{-\baselineskip}
\end{figure}

Interestingly, sources relating to previous diagnoses of thrombotic or lacunar stroke appeared to be highly protective against CS, despite the fact that occurrence of a first stroke is known to increase the risk of a subsequent stroke significantly across all types\cite{secondstroke}. This unexpected result is likely the unintended consequence of our sampling methods. Since the sampling window for all records labeled as positive only includes dates prior to the patient's first recorded stroke, we unintentionally blinded the model to instances of recurrent strokes. This likely caused the model to assume that such events were incredibly rare or impossible, leading to the conclusion that a prior stroke must somehow prevent future cryptogenic stroke. 

\section*{Discussion and Conclusions}

In this study, we proposed using machine learning algorithms to implement a probabilistic independence-based analysis in order to shed light into the relationship between cryptogenic stroke and migraine. 
We used ICA to disentangle 2000 independent latent sources of disease and their corresponding signatures from a large real-world EHR dataset of neurology patients. 
This was done unsupervised, which removes the inherent error and bias in human-driven data analysis, allowing for the data to speak for itself\cite{strobl,marco}. However, during signature discovery, we were limited by the memory constraints of our machine, forcing us to settle to untangle a maximum of 2000 signatures. 
The dimensional reduction during projection of our 9000-dimensional patient cross-sections onto the 2000-dimensional signature space implies that some information is lost, but to what extent is unclear. 
The natural implication is that multiple disease sources may be combined into one signature, reducing interpretability of our results. 
The results are also limited by the data available to the model; EHR data is inherently irregular and incomplete, with relevant clinical data never making its way into the record. Real world data is also notorious for introducing confounding into predictive models.
However, the large amount of records and the use of random forest models mitigate the impact of incomplete and noisy data \cite{marco,lasko2013,lasko}, and any potential confounding between sources is minimized by ensuring they are the algebraically independent error terms of the structural model and therefore the most downstream nodes on the causal graph\cite{marco,strobl,ica}. 

Our random forest model was internally validated on a held-out test set, and externally validated by verifying the observed causal effects of high-SHAP sources within the model with corresponding clinical evidence from the literature. 
The equivalence between the sources disentangled by ICA and the error terms in the patient-specific causal graph is what allows us to utilize SHAP values in this way to quantify the the causal effect each source has on each patient's final prediction within the model. This results in increased interpretability and usefulness beyond that seen with typical predictive models. Many clinically used risk calculators can identify patients at risk for a given disease, but cannot give reasons for \textit{why} that patient might be at risk. Therefore, causal models such as ours can be particularly useful for clinicians engaging in precision medicine. Even when the data is particularly sparse, incoherent, or complex, causal models can yield patient-specific results, allowing the interpreting clinician to give personalized explanations and recommendations for each patient, demonstrated by the waterfall plot in \textbf{Fig. 4}. 

On larger scale, this enhanced interpretability allows for the identification of the most common root causes of CS at a population level. Overall, the sources with the largest causative effects were protective and were characterized by pharmacological treatment with common preventative medications, lending validity to the model. However, two surprising drivers of cryptogenic stroke in migraine identified by the model were a source related to phenytoin, which was protective against CS, and a source related to allergic rhinitis, which was causative. 

While these results may indicate allergic rhinitis (AR) itself exerts a causal effect on ischemic stroke, a possibility that has been previously (though infrequently) discussed in the literature\cite{nose,nose2}, another plausible explanation is that this source represents the unobserved downstream effects of AR. It has previously been hypothesized that the chronic inflammation associated with AR could accelerate the development of atherosclerosis in the head and neck\cite{nose2}, leading to an increase in stroke risk. This could then be further exacerbated by the transient vasoconstriction seen in migraine episodes. Additionally, allergic rhinitis is frequently treated with over-the-counter medications\cite{otc} which are frequently not recorded in the EHR as they are often not reported by the patient\cite{pharm}. We suggest the possibility that this source may represent the cumulative effect of chronic usage of sympathomimetic decongestants such as oxymetazoline and pseudoephedrine, which have also been linked to CS in case reports and retrospective studies.\cite{sudafed,afrin} Testing of these hypotheses to properly characterize the link between allergic rhinitis and CS is needed and represents a promising direction for future investigation. 

\section*{Acknowledgements}
By using VUMC's Synthetic Derivative, this project was supported by CTSA award \#UL1TR000445 from the National Center for Advancing Translational Sciences (NCATS), a center of the U.S. National Institutes of Health (NIH). Its contents are the sole responsibility of the authors and do not represent official views of the NCATS or NIH.
\bibliographystyle{vancouver}
\bibliography{bibliography}

\end{document}